\documentclass[aps,pre,epsf,twocolumn,showpacs]{revtex4}
\UseRawInputEncoding
\usepackage{amsmath}
\usepackage{amssymb}
\usepackage{graphics}
\usepackage{graphicx}

\begin{document}
\title{Ising universality in the two-dimensional Blume-Capel model with quenched random crystal field}

\author{Erol Vatansever$^1$}
\author{Zeynep Demir Vatansever$^1$}
\author{Panagiotis E. Theodorakis$^2$}
\author{Nikolaos G. Fytas$^3$}
\altaffiliation[]{Corresponding author: nikolaos.fytas@coventry.ac.uk}

\affiliation{$^1$Department of Physics, Dokuz Eyl\"{u}l University, TR-35160, Izmir, Turkey}

\affiliation{$^2$Institute of Physics, Polish Academy of Sciences, Al.\
Lotnik\'ow 32/46, 02-668 Warsaw, Poland}

\affiliation{$^3$Centre for Fluid and Complex Systems, Coventry
University, Coventry, CV1 5FB, United Kingdom}

\date{\today}

\begin{abstract}
Using high-precision Monte-Carlo simulations based on a parallel version of the Wang-Landau algorithm and finite-size scaling techniques we study the effect of quenched disorder in the crystal-field coupling of the Blume-Capel model on the square lattice. We mainly focus on the part of the phase diagram where the pure model undergoes a continuous transition, known to fall into the universality class of the pure Ising ferromagnet. A dedicated scaling analysis reveals concrete evidence in favor of the strong universality hypothesis with the presence of additional logarithmic corrections in the scaling of the specific heat. Our results are in agreement with an early real-space renormalization-group study of the model as well as a very recent numerical work where quenched randomness was introduced in the energy exchange coupling. Finally,  by properly fine tuning the control parameters of the randomness distribution we also qualitatively investigate the part of the phase diagram where the pure model undergoes a first-order phase transition. For this region, preliminary evidence indicate a smoothening of the transition to second-order with the presence of strong scaling corrections.
\end{abstract}

 \maketitle

\section{Introduction}
\label{sec:introduction}

The effect of random disorder on phase transitions is one of the basic problems in
condensed-matter physics~\cite{young:book}. Examples include quantum Ising magnets
such as $\mathrm{LiHo_xY_{1-x}F_x}$ \cite{tabei2006,silevitch2007}, nematic liquid
crystals in porous media \cite{maritan1994}, noise in high-temperature
superconductors~\cite{carlson2006} and the anomalous Hall
effect~\cite{nagaosa2010}. Understanding random disorder in classical, equilibrium
systems is a crucial step towards solving the more involved problems in quantum
systems~\cite{vojta2014}, for example many-body localization with programmable random
disorder \cite{smith2016}, and in non-equilibrium phase
transitions~\cite{barghathi2012}.

The case of weak disorder coupled to the energy density of systems with continuous
transitions is relatively well understood: Uncorrelated disorder is relevant and leads to
new critical exponents if the specific-heat exponent $\alpha$ of the pure system is
positive, a rule known as the Harris criterion~\cite{harris74}. If long-range
correlations in the disorder are present, this rule can be generalized leading to
interesting ramifications~\cite{weinrib:83,luck:93a,wj:04a,barghathi:14,fricke:17,fricke:17b}.
These effects, and in particular the marginal case of a vanishing specific-heat
exponent as present in the two-dimensional (2D) Ising model, have
attracted a large research effort over the past decades~\cite{dotsenko81,dotsenko:83,
shalaev:84,shankar:87,ludwig:88,wang:90,selke:98,hasenbusch:08a,kenna:08,dotsenko:17}.

The situation is less clear for systems undergoing first-order phase transitions. 
The observation that formally $\nu = 1 / D$ and
$\alpha = 1$ for such systems in $D$ dimensions suggests that disorder is always
relevant in this case, and the general observation is that it indeed softens
transitions to become continuous~\cite{cardy:99a}. Such a rounding of discontinuities
has been rigorously established for systems in two dimensions~\cite{aizenman:89a},
but is believed to be more general -- a view that is supported by a mapping of the
problem onto the random-field model~\cite{hui:89a,berker93,cardy:97a}. 
Still, a number of important questions have
not been answered in full generality~\cite{cardy:97a,bellafard:12,zhu:15}, neither in two 
nor in three dimensions, where the main platform model was the random $q$-state Potts 
model~\cite{chen:92,picco:97,chatelain:01a,berche:03a,chatelain:05,delfino:17}. 

Another fertile testing platform for predictions relating to the universality 
principle of spin models under the influence of quenched disorder is the Blume-Capel
model~\cite{blume66,capel66}. This model has a long history and is linked to a diverse spectrum of 
actual experimental systems, including the prime nuclear fuel
uranium dioxide~\cite{blume66}, Mott insulators~\cite{kudin2002,lanata2017},
$^{3}$He--$^{4}$He mixtures~\cite{blume1971,lawrie84} and more general
multi-component fluids~\cite{wilding96}. The pure system features a tricritical point 
separating second- and first-order lines of transitions~\cite{silva06,malakis09,kwak15,zierenberg17} and it is well-known that several complications in the identification of criticality may arise under the presence of quenched disorder, as manifested by recent works on the topic~\cite{malakis10,theodorakis12,sumedha17,santos18}. Currently, the prevailing view is that the disorder-induced continuous transitions in both segments of the phase diagram of the model belong to the universality class of the pure Ising ferromagnet with logarithmic corrections, as shown in Ref.~\cite{fytas18}, where the random-bond Blume-Capel model has been investigated using high-precision numerical simulations. In fact, this appears to be the physically most plausible scenario given that both transitions are between the same ferromagnetic and paramagnetic phases (see also Fig.~\ref{fig:phase_diagram}), supporting the strong universality hypothesis~\cite{heuer1991,talapov1994,reis1996}. 

In the current work we provide additional evidence in favor of the strong universality hypothesis, by studying the Blume-Capel model but with a different type of quenched randomness in the crystal-field coupling parameter. A site-dependent crystal-field coupling has also been used in the past by Branco and Boechat~\cite{branco97} and more recently by Sumedha and Mukherjee~\cite{sumedha20} and is much closer to the experimental reality, as it mimics the physics of random porous media (mainly aerogels) in $^{3}$He--$^{4}$He mixtures~\cite{buzano94}. We employed extensive numerical simulations using a parallel implementation of the Wang-Landau algorithm~\cite{wang}, as outlined in the following Sec.~\ref{sec:model_methods}. The bulk of our simulations was performed in the original second-order transition regime of the pure system and our main result was reached not only by estimating the values of the standard critical exponents of the transitions, but also by inspecting the infinite-limit size extrapolation of the universal ratio $\xi/L$, where $\xi$ is the second-moment correlation length and $L$ the linear system size (see Sec.~\ref{sec:results}). Some additional preliminary results for the first-order transition regime of the phase diagram and for small disorder strength are also given at the end of Sec.~\ref{sec:results}, illustrating the softening of the transition but also the existence of strong scaling corrections. This contribution ends in Sec.~\ref{sec:conclusions} where a summary of our main findings alongside with an outlook for future work is given.  

\section{Model and Methods}
\label{sec:model_methods}

We consider the 2D spin-$1$ Blume-Capel model~\cite{blume66,capel66} as defined from the Hamiltonian
\begin{equation}\label{eq:Ham_dis}
  \mathcal{H}
  = -J\sum_{\langle xy \rangle}\sigma_{x}\sigma_{y}+\sum_{x}\Delta_{x}\sigma_{x}^{2} = E_{J}+E_{\Delta},
\end{equation}
where $J>0$ denotes the ferromagnetic exchange interaction coupling, the spin variables $\sigma_x\in\{-1,0,+1\}$ live on a square lattice with
periodic boundaries and $\langle xy\rangle$ indicates summation over nearest
neighbors. $\Delta_{x}$ represents the crystal-field strength and controls the density of vacancies ($\sigma_{x} = 0$). Following Ref.~\cite{branco97} and the experimental motivation~\cite{buzano94}, we choose in the current work a site-dependent bimodal crystal-field probability distribution of the form
\begin{equation} \label{eq:bimodal}
  \mathcal{P}(\Delta_{x})=p\delta(\Delta_{x}+\Delta)+(1-p)\delta(\Delta_{x}-\Delta),
\end{equation}
where $p\in (0,1)$ is the control parameter of the disorder distribution.

\begin{figure}
\includegraphics[width=85mm]{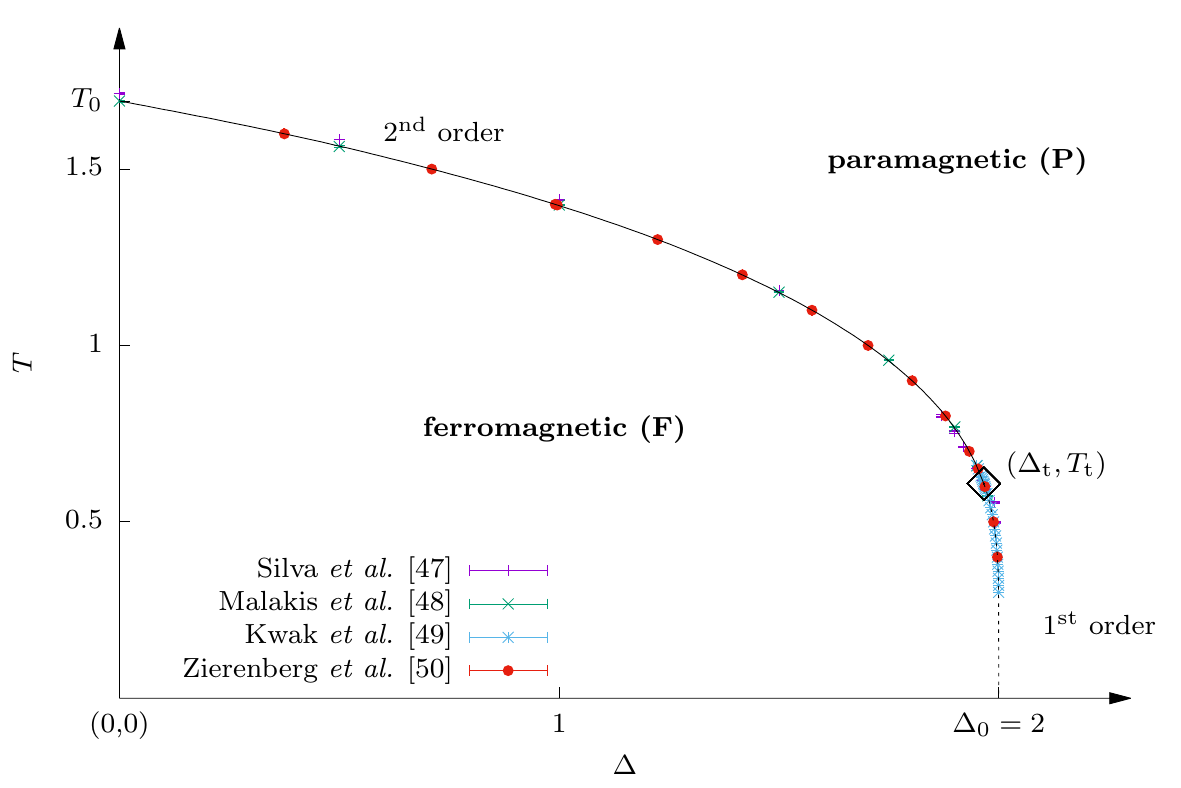}
\caption{Phase diagram of the pure ($p=0$) 2D
Blume-Capel model in the crystal-field -- temperature plane~\cite{zierenberg17}, showing the ferromagnetic
(\textbf{F}) and para\-magnetic (\textbf{P}) phases that are
separated by a continuous transition for small $\Delta$ (solid
line) and a first-order transition for large $\Delta$ (dotted
line). The line segments meet at a tricritical point ($\Delta_{\rm t}$, $T_{\rm t}$), as indicated
by the black rhombus. The data shown are selected estimates from previous studies.  \label{fig:phase_diagram}}
\end{figure}

For $\Delta = \infty $ the model is equivalent to the random site spin-$1/2$ Ising model, where sites are present or absent with probability $p$ or $1-p$, respectively~\cite{branco97}. This comes from the fact that, for $\Delta = \infty$, a $+\Delta$ crystal field acting on a given site $x$ forces that site to be in the $\sigma_{x} = 0$ state, while a $-\Delta$ crystal field forces the site to be either in the state $\sigma_{x} = +1$ or in the state $\sigma_{x} = -1$. Thus, only for high enough $p$ an infinite cluster of $\sigma_{x} = \pm 1$ states will form and will be able to sustain order. Exactly at $p = p_{\rm c}$, there is such an infinite cluster but its critical temperature is zero. In Ref.~\cite{branco97} $p_{\rm c}$ has been estimated to be $0.5$ and not $0.5927$ as expected for the site percolation problem~\cite{stauffer}. This discrepancy was attributed to the nature of the small-cell real-space renormalization-group method used.

For $p = 0$ the pure Blume-Capel model is recovered (for a review see Ref.~\cite{zierenberg17}). The phase diagram of the pure model in the ($\Delta$, $T$)--plane is shown
in Fig.~\ref{fig:phase_diagram}: For small $\Delta$ there is a line of continuous transitions (in the Ising universality class)
between the ferromagnetic and paramagnetic phases that crosses the $\Delta = 0$ axis
at $T_{0}\approx 1.693$~\cite{malakis10}. For large $\Delta$ the
transition becomes discontinuous and it meets the $T=0$ line at
$\Delta_0 = zJ/2$~\cite{capel66}, where $z = 4$ is the coordination number (here we
set $J = 1$, and also $k_{\rm B} = 1$, to fix the temperature scale). The two line
segments meet in a tricritical point estimated to be at
$(\Delta_{\rm t}\approx1.966,T_{\rm t}\approx0.608)$~\cite{kwak15,jung17}. The crucial observation here is that with the inclusion of disorder ($p > 0$), it is expected that the value of $\Delta_0$ will increase. This can be clearly seen from the results of Ref.~\cite{branco97}, where for $p = 0.1$ one obtains $\Delta_0 \approx 2.2$, for $p=0.3$, $\Delta_0 \approx 3.5$, and eventually $\Delta_0$ diverges for $p = p_{\rm c}$. For the bulk simulations of the current work the control parameter $p$ was set to the value $p = 0.5$ (unless otherwise stated) and, as it will be seen below, the obtained results are fully consistent with this renormalization-group prediction.

\begin{figure}
\includegraphics[width=95mm]{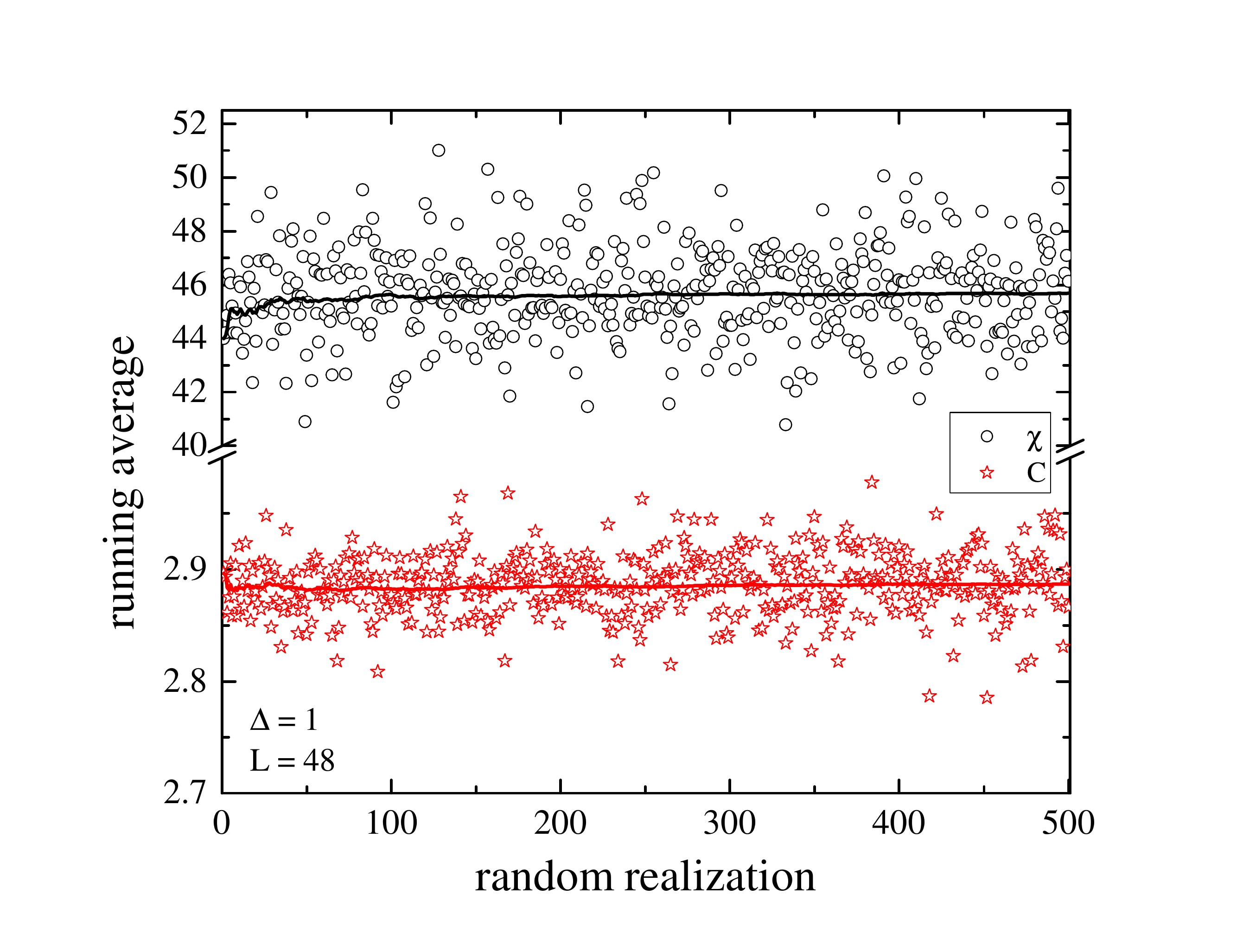}
\caption{Disorder distribution of the susceptibility $\chi$ and the specific heat $C$ maxima for a system with linear size $L = 48$, where $p=0.5$ and $\Delta = 1$. The running averages over the samples are shown by the solid lines. \label{fig:run_average}}
\end{figure}

On the other hand, the expected effect of any small disorder $0 < p \ll 0.5$
in the original first-order transition line would be to either soften all to
a continuous transition at once, see Ref.~\cite{branco97}, or to decrease its extent 
continuously with $p$ for temperatures below $T_{\rm t}$ 
up to a certain value $p^{\ast}$, above which there is no first-order
transition. This was observed in the recent work of Ref.~\cite{sumedha20} for the present model in a fully connected graph, where $p^{\ast} \approx 0.1$ was found. Irrespective of the underlying graph topology of these analytical results,
the effect of disorder in the first-order transition regime of the pure model can therefore only be addressed in transitions occurring in the vicinity of the tricritical point or for the $T < T_{\rm t}$ temperature range, or
even in a more controlled set of parameters for simulations in the small $p$-limit (say for $p \leq ~0.1$). In view of this interesting observations we also qualitatively probed this regime of the phase diagram in order to locate signatures of first-order transition and any smoothening effects due to the presence of quenched randomness.

\begin{figure}
\includegraphics[width=95mm]{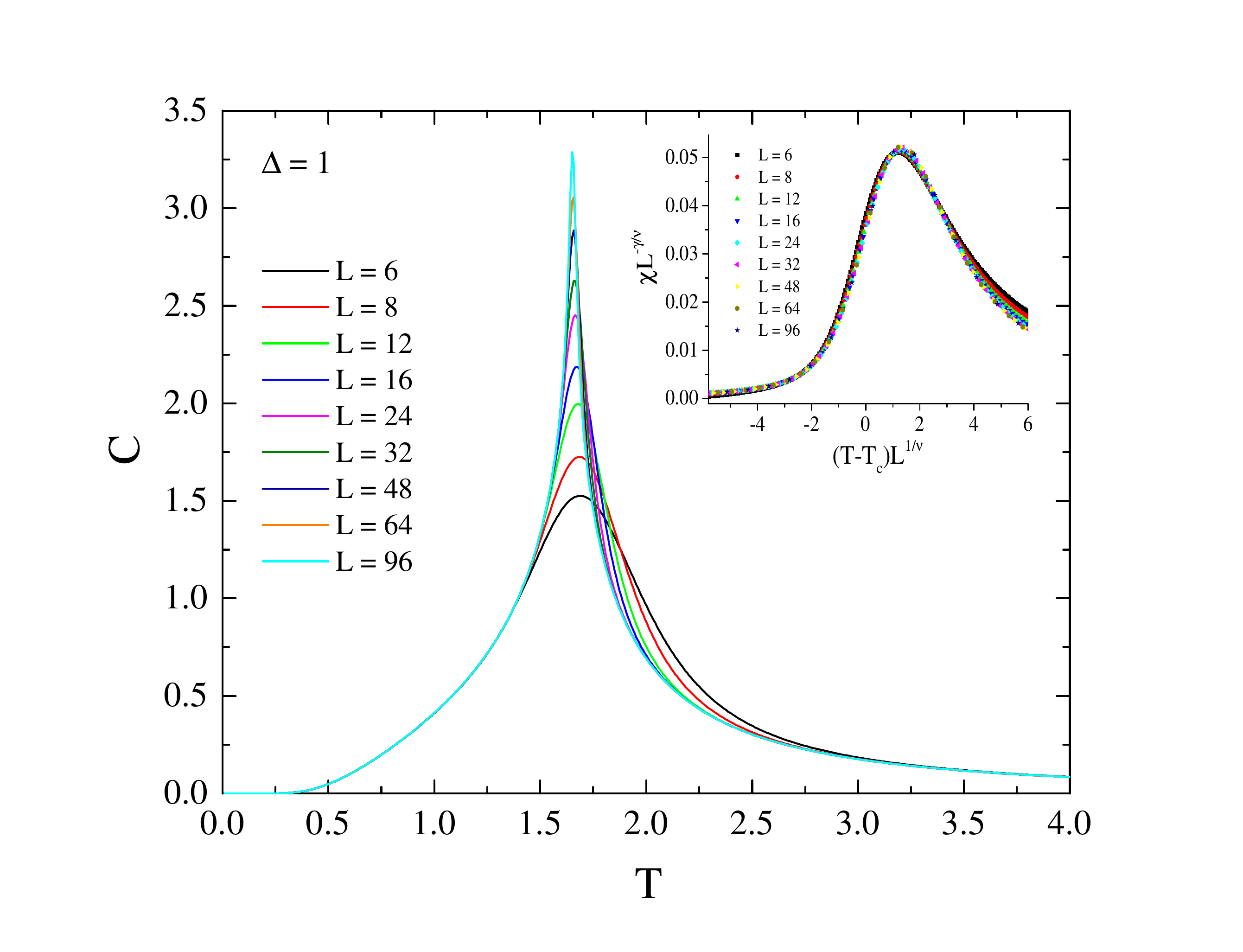}
\caption{{\bf Main panel}: Specific-heat curves as a function of the temperature for the whole range of system sizes studied for $\Delta = 1$ and $p = 0.5$. Note the clear signs of a shift behavior of the peak characteristic of a second-order phase transition. {\bf Inset}: The corresponding susceptibility data for the same set of parameters displayed using the data collapse method.
  \label{fig:curves}}
\end{figure}

To study the interesting phenomena discussed above we employed Monte Carlo simulations and in particular the well-known Wang-Landau algorithm~\cite{wang}. This algorithm is a valid choice for the present work as the model under study is not expected to have any replica-symmetry breaking -- note that the ergodic hypothesis is equivalent to the absence of replica-symmetry breaking~\cite{sommers83}. In a Wang-Landau simulation random walks are performed in the energy space and trial spin configurations are accepted with a probability proportional to the reciprocal of the density of states $g(E)$. During the simulation, the energy histogram is also accumulated. If the histogram is ``flat'' (requirement that the number of visits at each energy level is not less than $x\%$ of the average histogram), the density of states is then modified by a multiplicative factor $f$ and the new density of states is used in the next random walk. The final density of states is then obtained at the end of the simulation and does not depend on the temperature. Therefore, it is possible to compute the full spectrum of the thermodynamic quantities of interest such as the energy $E$ and the magnetization $M$ (including their variance-related quantities: the specific heat $C$ and susceptibility $\chi$) at any temperature with a single run. 

In special cases like the present one, where the energy of the system can be split into two parts, one may also focus on the computation of the joint density of states, $g(E_{J},E_{\Delta})$, that provides access to the wider phase space of the system. However, Wang-Landau simulations for multi-energy variables cost significantly more in computational time than their one-variable counterpart, and this is the reason why the simulations of Ref.~\cite{kwak15} for the pure 2D Blume-Capel model were upper-bounded by sizes of $L\le 48$. In the present work and in order to obtain access to larger system sizes, we chose to accumulate the 1D density of states, with the cost of repeating the simulations at different values of $\Delta$.

\begin{figure}
\includegraphics[width=95mm]{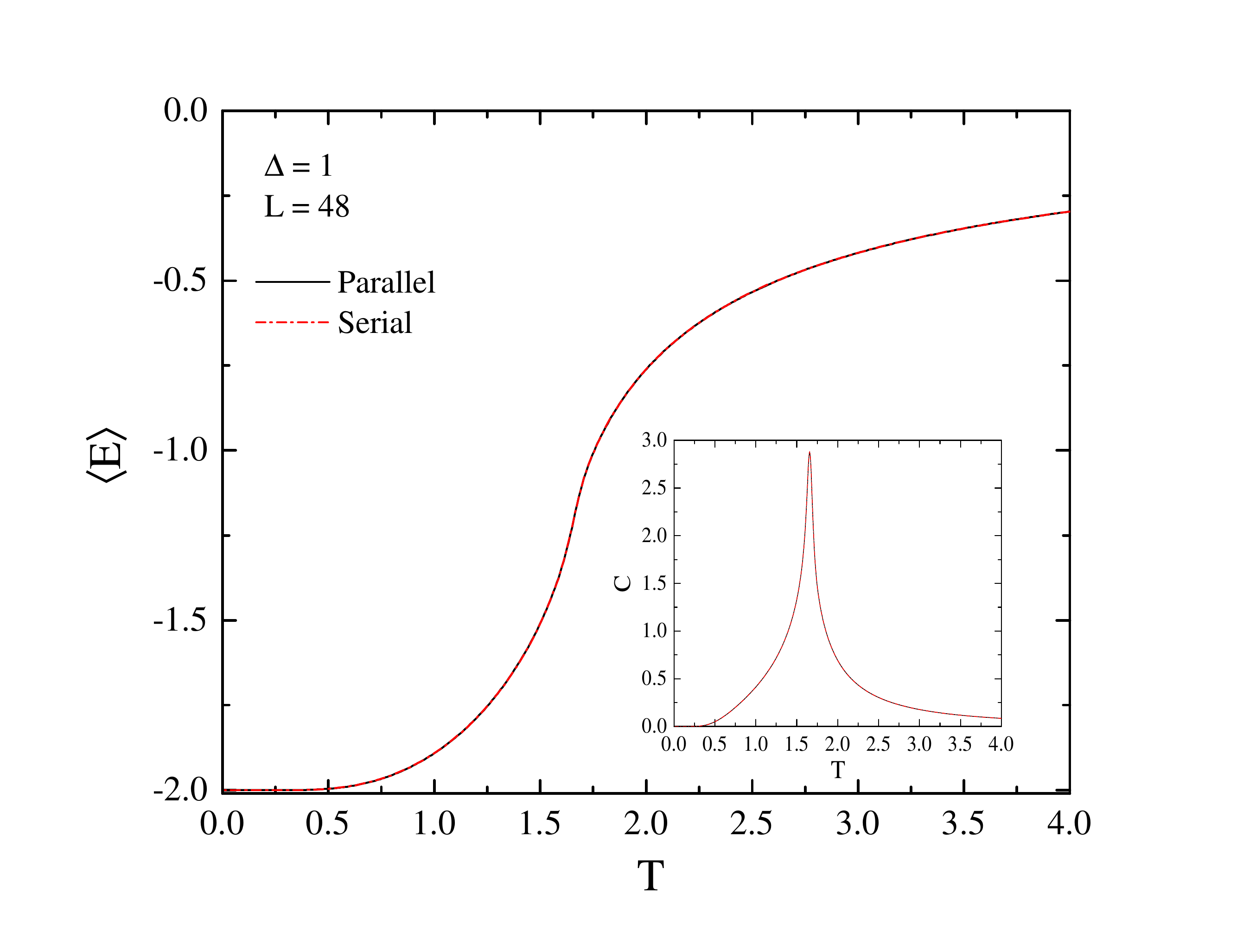}
\caption{Energy ({\bf main panel}) and specific-heat ({\bf inset}) curves as a function of the temperature for a system with linear size $L=48$, for $\Delta = 1$ and $p = 0.5$, by using a serial and a parallel version of the Wang-Landau algorithm. The numerical data of the parallel version have been obtained using 28 processes on equivalent CPU cores. \label{fig:numerics}}
\end{figure}

To date, many efforts have been recorded in the relevant literature with respect to understanding and further improving the performance of the Wang-Landau algorithm~\cite{zhou05}. One line of research refers to the  parallelization of the algorithm in two different directions: (i) Dividing the total energy space into smaller subspaces, each then being sampled by independent random walkers~\cite{wang,vogel}, and (ii) Several random walkers work simultaneously on the same density of states by using distributed memory~\cite{khan}, shared memory~\cite{zhan}, and graphics processing unit~\cite{yin} architecture. In the present work, we implemented the latter scheme and proposed a distributed memory implementation of the Wang-Landau algorithm using Message-Passing Interface  architecture.

Our parallel implementation performs the following steps:
\begin{enumerate}

 \item Every processor generates its starting configuration and corresponding initial energy $E_0$ using different random seeds. In the beginning of the simulation, the modification factor is set to $\ln{(f)} =1$ for every processor.
  
 \item The density of states and the histogram of every processor are initialized follows: $g(E)=1$ and $H(E)=0$.
 
 \item All the processors involved in the simulation perform standard Wang-Landau procedure as follows:
 A trial move is made by randomly selecting a spin. If $E_{\rm old}$ and $E_{\rm new}$ are the energies before and after the proposed move, respectively, the transition to new state is accepted via the Wang-Landau probability
 \begin{equation}
 \Pi(E_{\rm old}\rightarrow E_{\rm new})=\min[1,g(E_{\rm old})/g(E_{\rm new})].\nonumber
  \end{equation}
If the trial state is accepted, the histogram is then updated as $H(E_{\rm new})\rightarrow H(E_{\rm new})+1$ and the corresponding density of states is multiplied by the modification factor: $g(E_{\rm new}) \rightarrow f \cdot g(E_{\rm new})$. If the proposed move is not accepted, the existing histogram is modified  as $H(E_{\rm old}) \rightarrow  H(E_{\rm old})+1$ and the density of states is updated as $g(E_{\rm old}) \rightarrow f \cdot g(E_{\rm old})$. The Wang-Landau procedure is applied until one Monte Carlo sweep is completed -- one sweep is $N$ spin-flip attempts, where $N = L^{2}$ denotes the total number of spins on the lattice. 

\item After the completion of a Monte Carlo sweep, each processor sends its density of states and the relevant histogram to the master processor. Then, the master processor combines the inputs in order to calculate the global density of states and energy histogram. Then, the global density of states are redistributed to all processors. 
 
\item The master processor also checks whether the flatness is satisfied or not after every $10^4$ Monte Carlo sweeps. If the flatness criterion has been achieved, the modification factor is reduced as $f_{i+1} = \sqrt{f_i}$. Then, the histogram of every processor is set to $H(E) = 0$ and steps 3 and 4 are repeated for the new modification factor. 

\end{enumerate}

\begin{figure}
\includegraphics[width=95mm]{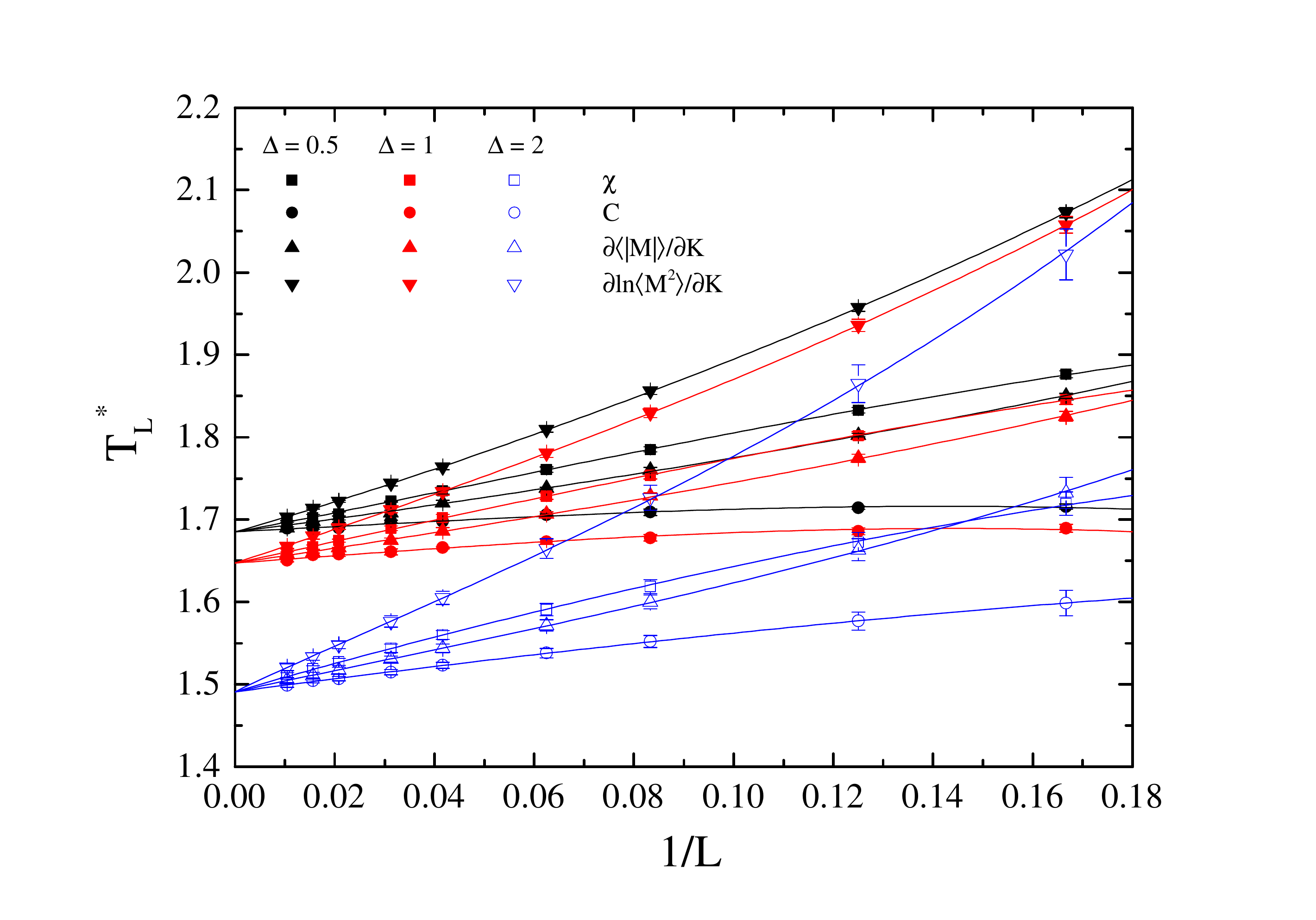}
\caption{Shift behavior of several pseudo-critical temperatures defined in the text as a function of the inverse system size for $\Delta = 0.5$, $1$, and $\Delta= 2$, where $p = 0.5$ in all cases. \label{fig:pseudo_T}}
\end{figure}

For small system sizes, the above scheme has the disadvantage that the time required for the communication among the processors may surpass the time needed for the actual computation. Nevertheless, the method works predominantly well with increasing system size, allowing us in the current work to simulate efficiently linear sizes up to $L = 96$. The flatness criterion for the histogram used was $80\%$ for all the lattice sizes considered. For the stopping criterion we used $\ln{(f_{\rm final})}=10^{-8}$ for $L < 96$ and $\ln{(f_{\rm final})}=10^{-7}$ for $L = 96$. 

We simulated the model defined in Eqs.~(\ref{eq:Ham_dis}) and (\ref{eq:bimodal}) at three values of the crystal-field coupling, namely $\Delta=0$, $1$, and $\Delta = 2$, fixing the control parameter $p$ to the value $0.5$.
For each value of $\Delta$ we considered the following sequence of linear sizes $L\in \{6, 8, 12, 16, 24, 32, 48, 64, 96\}$ and for each pair ($L$, $\Delta$) disorder averaging has been performed over $500$ random realizations -- see the characteristic running-average tests of Fig.~\ref{fig:run_average}. For the particular case of $\Delta = 2$ we also varied the parameter $p$ in the regime $0 < p \leq 0.1$ using sizes up to $L = 48$ and a moderate disorder averaging over $200$ samples. 

From this point and on all the figures shown below correspond to the disorder-averaged data, unless otherwise stated. Some typical specific-heat and susceptibility curves are shown in Fig.~\ref{fig:curves} for the case $\Delta = 1$ and for $p=0.5$. In particular, in the inset of this figure the susceptibility data are plotted using the data collapse method, which allows the reader to judge directly how well the data for different sizes are
approximated by the finite-size critical scaling function. Here, we have used the Ising values $\nu = 1$ and $\gamma/\nu = 1.75$ for purely illustrative reasons and our estimate $T_{\rm c} = 1.6473$, see Fig.~\ref{eq:pseudo_T} below. As it is clear from the plot, the smaller sizes $L\leq 16$ slightly deviate from the collapse due to the presence of the well-known finite-size effects. These effects have been taken into account in an effective way via the scaling ansatz used below and the corrections-to-scaling exponent $\omega$.

\begin{figure}
\includegraphics[width=95mm]{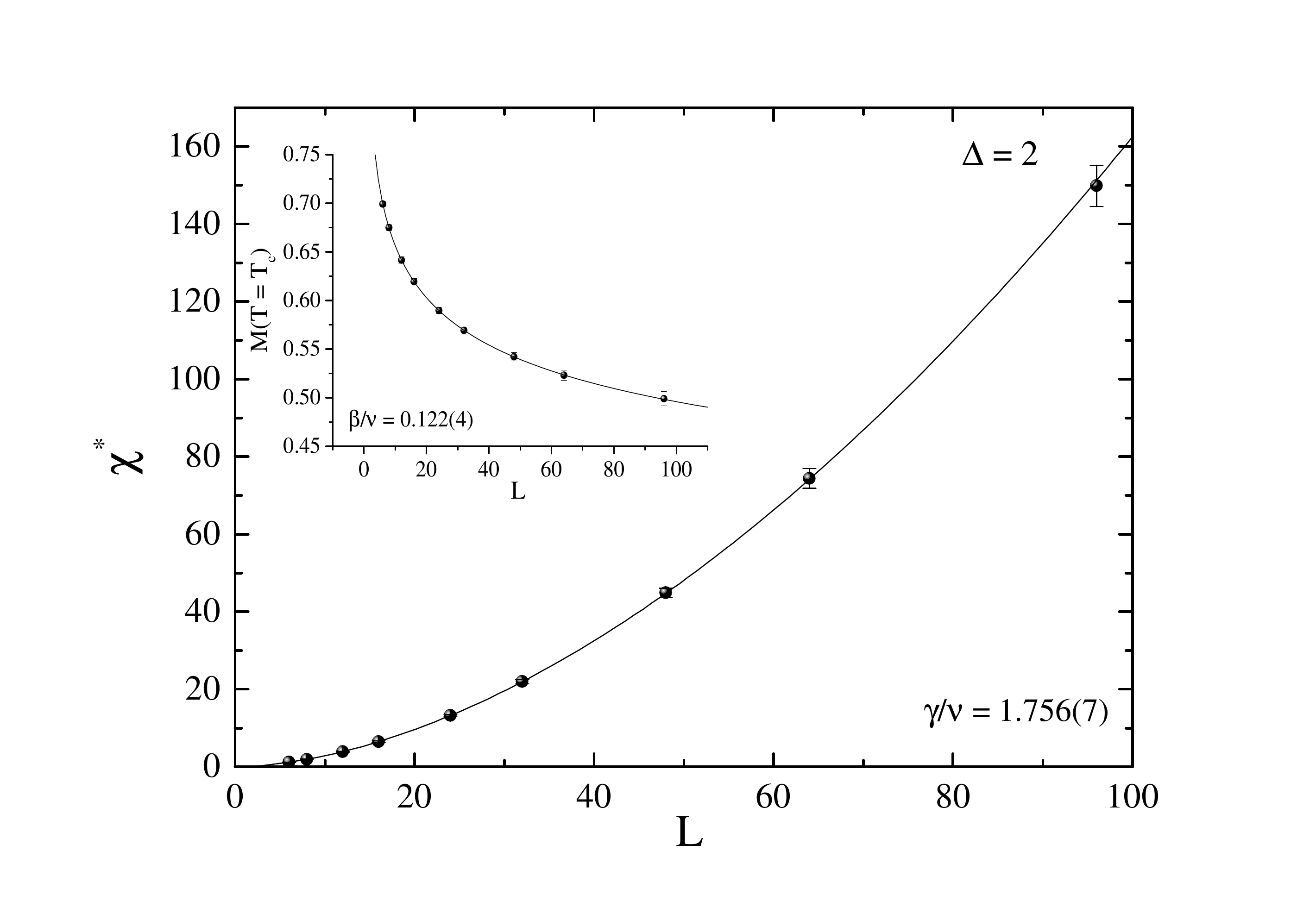}
\caption{Estimation of the magnetic exponent ratios $\gamma/\nu$ ({\bf main panel}) and $\beta/\nu$ ({\bf inset}) for the case $\Delta = 2$ and for $p = 0.5$, via the finite-size scaling behavior of the susceptibility maxima ($\chi^{\ast}$) and the order parameter at the estimated critical temperature. \label{fig:chi_mag}}
\end{figure}

Finally, statistical errors have been estimated using the standard jackknife method~\cite{barkema} and for the fitting procedure discussed below in cases needed we restricted ourselves to data with $L 􏰚\geq L_{\rm min}$. As usual, to determine an acceptable $L_{\rm min}$ we employed the standard $\chi^{2}$ test for goodness of fit. Specifically, the $p$ value of our $\chi^2$ test -- also known as $Q$, see, \emph{e.g.}, Ref.~\cite{press} -- is the probability of finding a $\chi^{2}$ value which is even larger than the one actually found from our data. Recall that this probability is computed by assuming Gaussian statistics and the correctness of the fit's functional form. We consider a fit as being fair only if $10\% < Q < 90\%$.
 
Before proceeding with the finite-size scaling analysis in the following Sec.~\ref{sec:results} we would like to underline that, at least to our knowledge, a parallel implementation of the Wang-Landau algorithm as described above has not been yet attempted in the literature of classical spin models. Thus, one additional goal of our study was to show the accuracy of the numerical scheme. For this purpose, we present in Fig.~\ref{fig:numerics} a comparative plot of the parallel and serial simulation results of the energy $E$ (main panel) and specific heat $C$ (inset) as a function of the temperature for a linear size $L = 48$ and crystal-field coupling $\Delta = 1$ with $p=0.5$. These curves clearly show that the numerical data from the two different implementations are practically indistinguishable. For the sake of completeness we have checked all the remaining thermodynamic quantities as well and similar behavior was observed,  but is not shown here for reasons of brevity. We have also performed extensive test runs at various lattice sizes and crystal-field strengths. Based on these tests we concluded that the number of CPU's used in the computation did not affect the numerical accuracy of our results. 
 
\section{Results}
\label{sec:results}

\begin{figure}
\includegraphics[width=95mm]{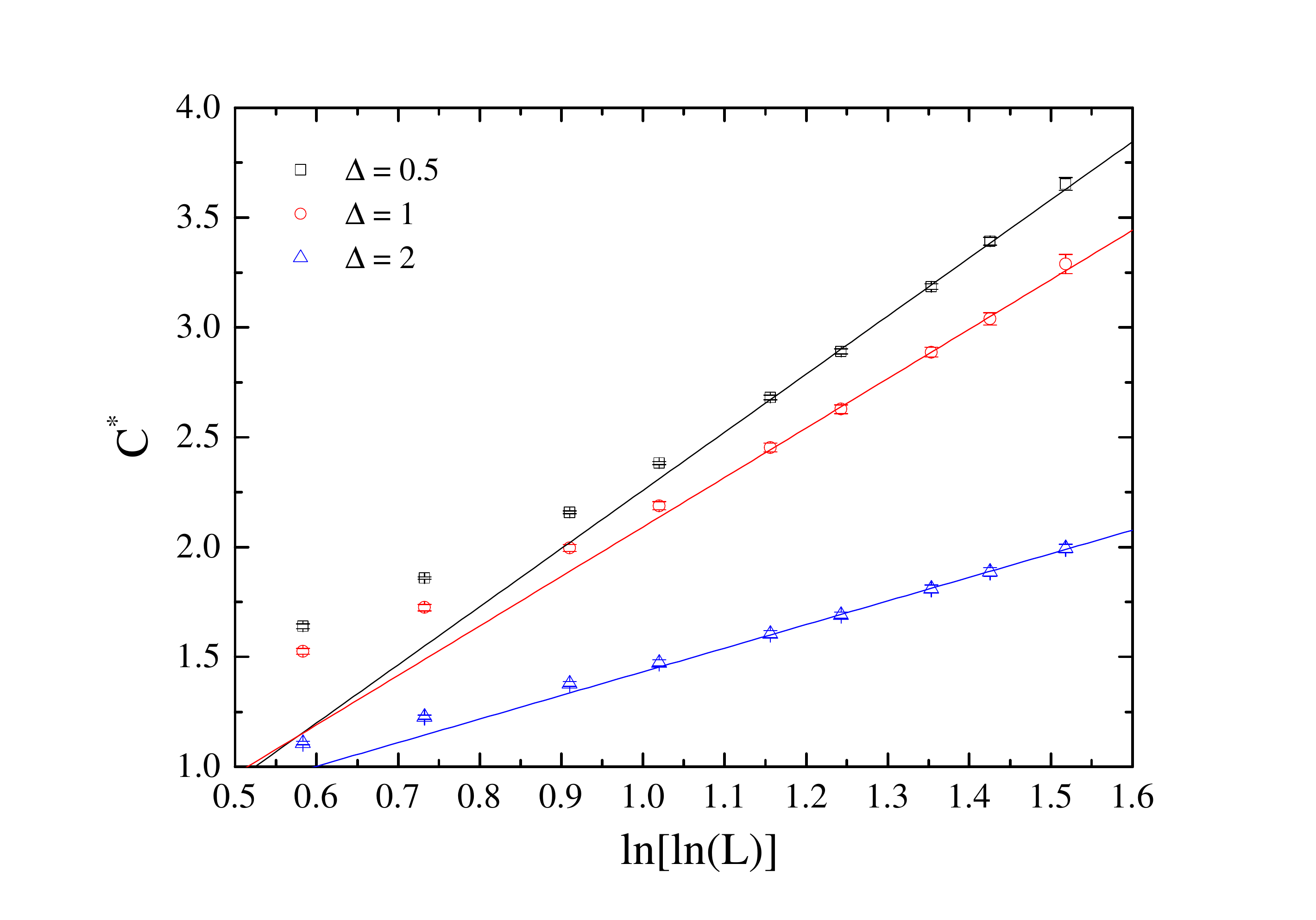}
\caption{ Finite-size scaling behavior of the specific-heat maxima for all values of $\Delta$ considered and for $p = 0.5$. The data are shown as a function of the double logarithm of the system size $L$. The solid lines are excellent linear fits using only the larger system sizes $L \geq L_{\rm min} = 24$. \label{fig:spec_heat}}
\end{figure}

We start the presentation of our results with Fig.~\ref{fig:pseudo_T} where we show the shift behavior of several pseudo-critical temperatures, $T_{L}^{\ast}$, corresponding to the maxima of the susceptibility $\chi$, the specific heat $C$, the derivative of the absolute order parameter $M$ with respect to inverse temperature ($K=1/T$), $\partial \langle |M|\rangle / \partial
K=\langle |M|H\rangle-\langle |M|\rangle\langle H\rangle$, and the logarithmic derivative of the second power of the order parameter with respect to $K$, $\partial \ln \langle M^{2}\rangle / \partial
K = \langle M^{2}H\rangle / \langle M^{2}\rangle-\langle
H\rangle$~\cite{ferrenberg91}. Three data sets are shown for the three values of $\Delta$ considered. Lines of different colors correspond to separate joint fits for each value of $\Delta$ with a common extrapolation to the expected power-law behavior
\begin{equation}\label{eq:pseudo_T}
T^{\ast}_{L}=T_{c}+bL^{-1/\nu}(1+b'L^{-\omega}).
\end{equation}
In the above equation $T_{\rm c}$ denotes the critical temperature and is $\Delta$-dependent, whereas $\nu$ and $\omega$ are universal critical exponents. In particular $\nu$ is the critical exponent of the correlation length and $\omega$ the corrections-to-scaling exponent, which was fixed to the 2D Ising universality class value $7/4$ (see also in Fig.~\ref{fig:chi_mag} below)~\cite{blote,shao}. The results for the critical temperatures obtained from the fits shown in Fig.~\ref{fig:pseudo_T} are as follows: $T_{\rm c}(\Delta = 0.5) = 1.6854(9)$, $T_{\rm c}(\Delta = 1) = 1.6473(7)$, and $T_{\rm c}(\Delta = 2) = 1.4907(6)$. The critical exponent $\nu$ was estimated to be $\nu(\Delta = 0.5) = 0.95(6)$,  $\nu(\Delta = 1) = 0.99(4)$, and $\nu(\Delta = 2) = 1.04(5)$. 

\begin{figure}
\includegraphics[width=95mm]{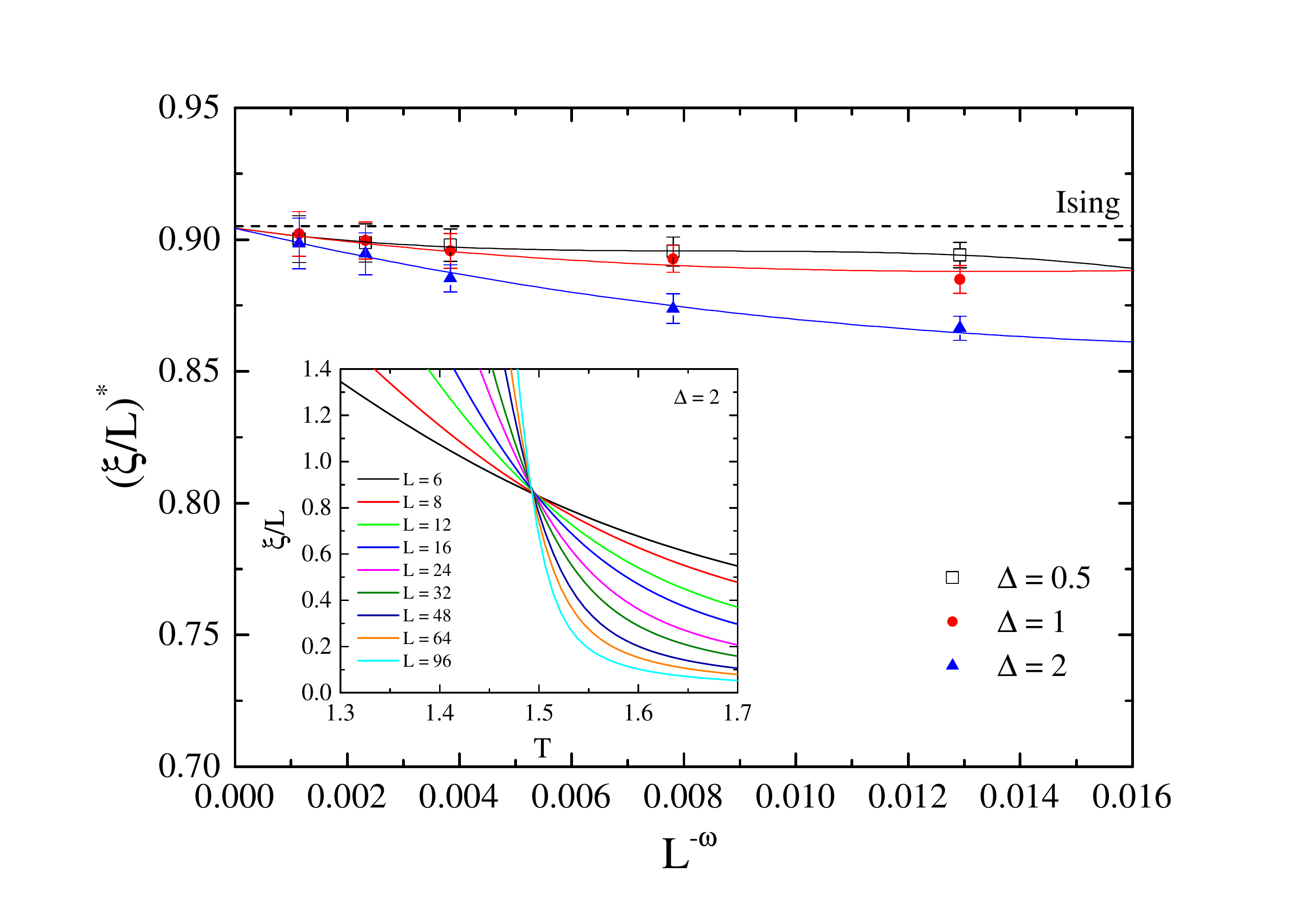}
\caption{{\bf Main panel}: Finite-size scaling of the correlation-length ratios at their crossing points, $(\xi/L)^{\ast}$, for the square-lattice Blume-Capel model with a quenched random crystal field. Results are shown for all three values of $\Delta$ and for the following pairs $(L,2L)$ of system sizes: $(12,24)$, $(16,32)$, $(24,48)$, $(32,64)$, and $(48,96)$. As in all previous plots, $ p =0.5$. The horizontal dashed line shows the asymptotic value for the square-lattice Ising model with periodic boundaries, see Eq.~(\ref{eq:ratio-exact}). The solid lines show joint polynomial fits of third order in $L^{-\omega}$ with a common extrapolation. {\bf Inset}: Typical $\xi/L$ curves as a function of the temperature for various system sizes for the case $\Delta = 2$. The temperature area of the crossings conforms to the value $T_{\rm c}(\Delta = 2) \approx 1.49$ of Fig.~\ref{fig:pseudo_T}. \label{fig:cor_length}}
\end{figure}

A few comments are now in order:

\begin{itemize}

\item For the values $\Delta = 0.5$ and $1$ we observe only a slight decrease in the critical temperature with increasing $\Delta$ and only for $\Delta = 2$ a downward trend of the critical temperatures starts to settle in. This is consistent with the qualitative results shown in Fig. 4 of Ref.~\cite{branco97}.

\item The values for the critical temperatures of the disordered model appear to be higher that those of the pure model (see Fig.~\ref{fig:phase_diagram} for a comparison), especially for the case $\Delta = 2$, where 
the critical temperature rises from $T_{\rm c} = 0 \rightarrow 1.4907$. This is also in full agreement with the results presented in Figs.~3 and 4 of Ref.~\cite{branco97}. A simple argument supporting this observed increase in the critical temperature is as follows: The case $p=0$ corresponds to the pure model for which all crystal fields are $+\Delta$, whereas the $p = 0.5$ case brings to the model  $-\Delta$ crystal fields which in turn favor the $\pm 1$ states. 

\item A similar enhanced ferromagnetic ordering has been also highlighted in the scaling analysis of the 2D random-bond Blume-Capel model~\cite{malakis09}. 

\item All our estimates for the critical exponent $\nu$ are compatible within error bars to the value $\nu=1$ of the 2D Ising universality class. 

\end{itemize}

\begin{figure}
\includegraphics[width=95mm]{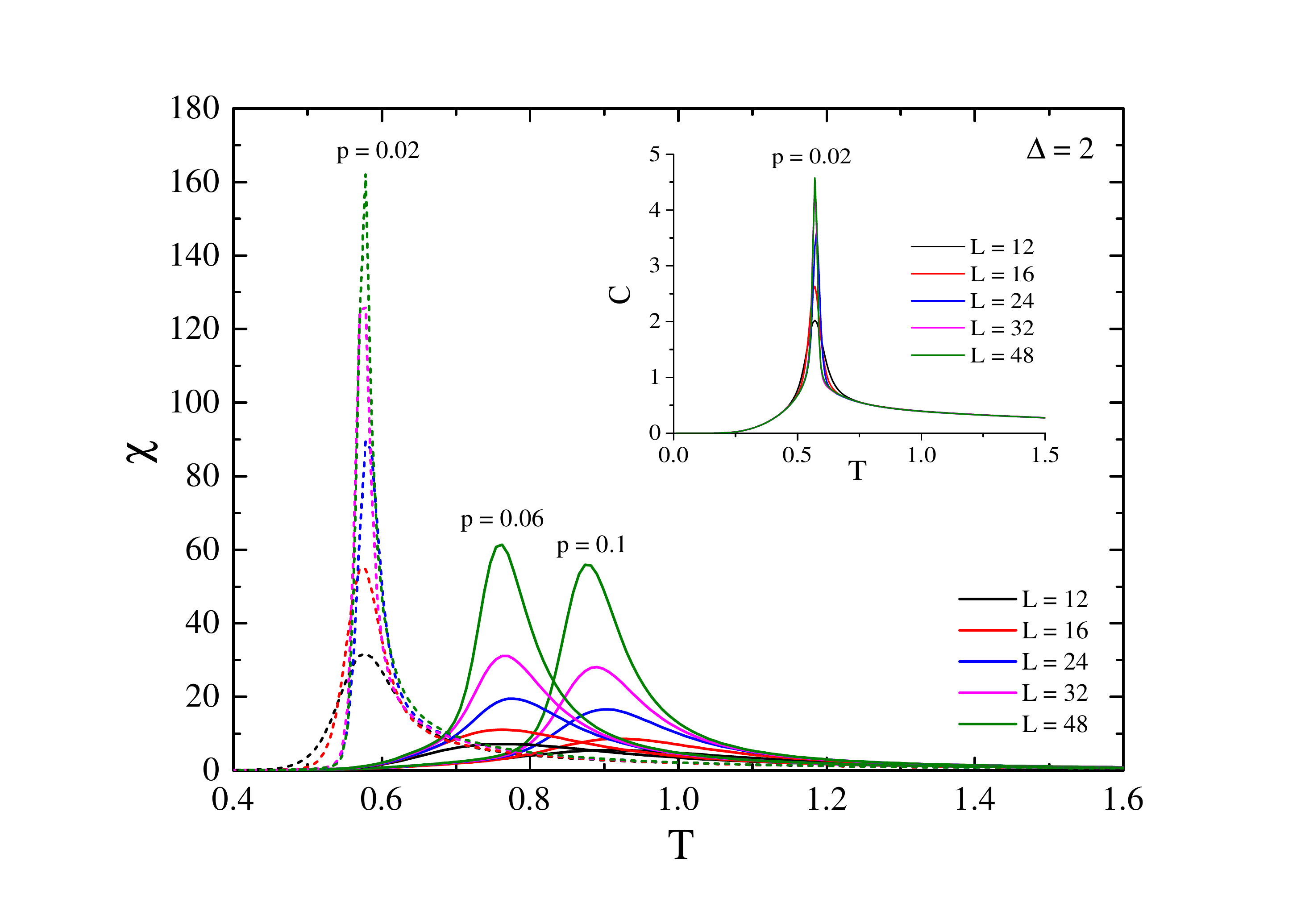}
\caption{{\bf Main panel}: Susceptibility curves as a function of the temperature for $p=0.1$, $0.06$ (solid lines) and $p=0.02$ (dashed lines). {\bf Inset}: Specific-heat curves as a function of the temperature for $p=0.02$. In both panels $\Delta = 2$.
  \label{fig:chi_small_p}}
\end{figure}

Additional evidence in this respect is given in Fig.~\ref{fig:chi_mag} where the finite-size scaling behavior of the susceptibility maxima $\chi^{\ast}$ (main panel) and the order parameter at the critical point (inset) are shown as a function of the system size for the case $\Delta = 2$. The solid lines are fits of the form $\chi^{\ast} \sim bL^{\gamma/\nu}(1+b'L^{-\omega})$ and $M(T = T_{\rm c}) \sim L^{-\beta/\nu}(1+b'L^{-\omega})$, respectively. The estimated values of the magnetic exponent ratios are: $\gamma/\nu  = 1.756(7)$ and $\beta/\nu = 0.122(4)$, both in agreement with the 2D Ising universality values $\gamma/\nu = 7/4$ and $\beta/\nu = 1/8$. Similar results have been also obtained for the cases $\Delta = 0.5$ and $1$, but are omitted here for brevity. However, we find it useful for the reader to quote the fitting results: $[\gamma/\nu, \beta/\nu] = [1.754(5), 0.124(3)]$ for $\Delta = 0.5$ and $[1.756(6), 0.126(3)]$ for $\Delta = 1$.

All our results up to this point support the strong universality hypothesis according to which the effect of infinitesimal disorder gives rise to a marginal irrelevance of randomness and besides logarithmic corrections, the critical exponents maintain their 2D Ising values. One of the most interesting observables in this framework is the specific heat that is expected to slowly diverge with a double-logarithmic dependence of the form 
\begin{equation}\label{eq:spec_heat}
C^{\ast} \sim \ln{[\ln{(L)}]}.
\end{equation}
We have tested successfully this scaling ansatz for all the available numerical data of the current work -- see Fig.~\ref{fig:spec_heat} and the relevant discussion in the caption.

\begin{figure}
\includegraphics[width=95mm]{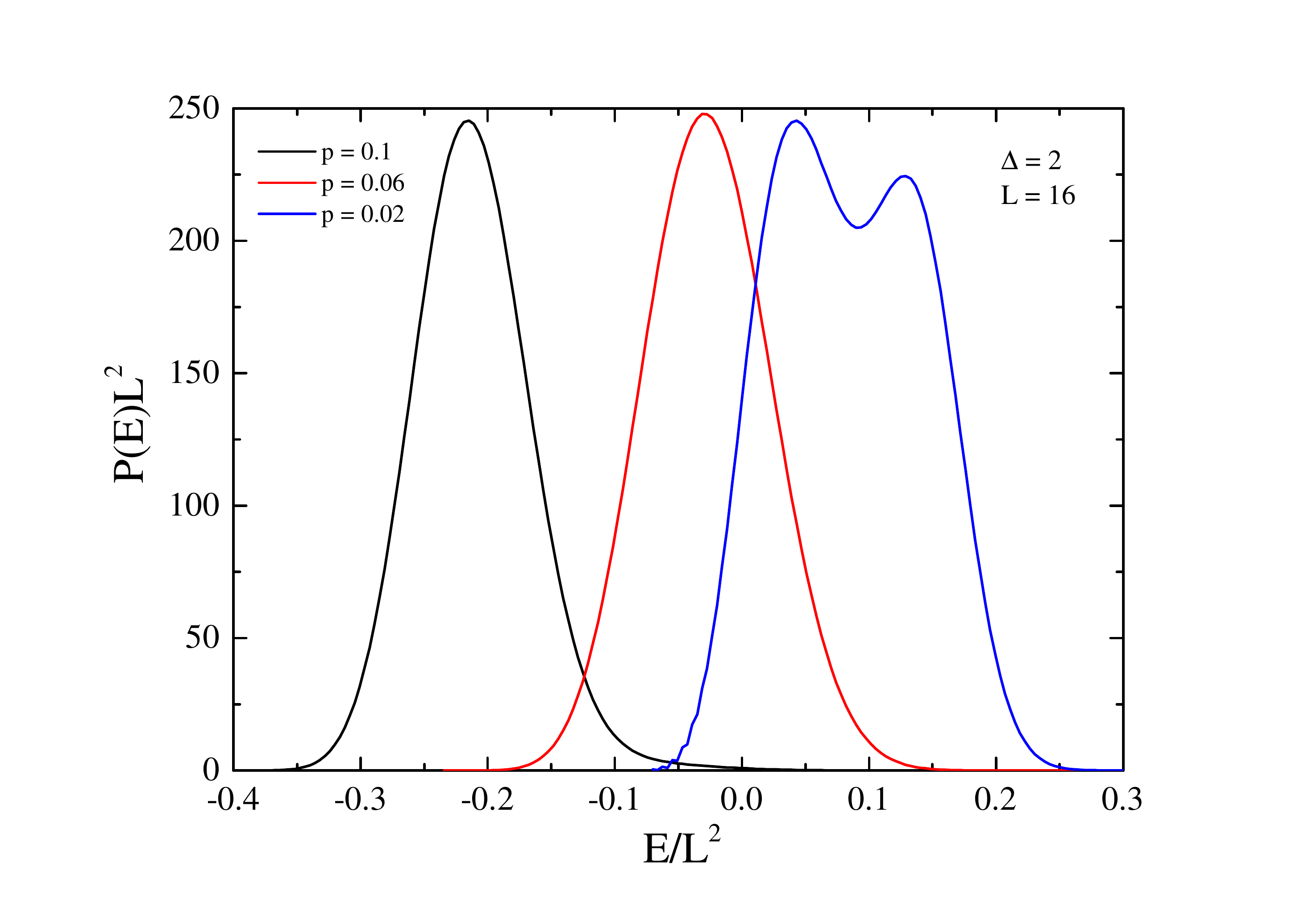}
\caption{Energy probability density functions of a system with linear size $L = 16$ at $\Delta = 2$ for $p=0.1$, $0.06$ and $p=0.02$. The curves have been computed at the pseudocritical points of the specific heat.
  \label{fig:energy_pdf_L16}}
\end{figure}

While universality classes are characterized by the entirety of their critical exponents, other useful universal amplitudes do exist and constitute additional strong evidence, allowing in certain cases to monitor non-monotonic scaling behavior in the approach to the thermodynamic limit (see Refs.~\cite{fytas18,fytasRFIM}). Such universal ratios are the well-known Binder cumulant but also the ratio $\xi/L$. In the current work we studied in detail the ratio $\xi/L$ and its size evolution for all range of parameters considered. This is known to be universal for a given choice of boundary conditions and aspect ratio. For Ising spins on a square lattice with periodic boundary conditions as $L \to \infty$
it approaches the value~\cite{salas2000}
\begin{equation} \label{eq:ratio-exact}
  \left(\frac{\xi}{L}\right)_{\infty, \; \rm pure} = 0.905\,048\,829\,2(4).
\end{equation}
It worth noting that the behavior of the pure and random-bond square-lattice Blume-Capel model was found to be perfectly consistent with Eq.~(\ref{eq:ratio-exact})~\cite{zierenberg17,fytas18}.

Following Ref.~\cite{fytas18}, for the estimation of $\xi$ we used its second-moment definition~\cite{cooper1989, ballesteros2001}: From the Fourier
transform of the spin field, $\hat{\sigma}(\mathbf{k}) = \sum_{\bf
x}\sigma_{\bf x}\exp(i{\bf kx})$, we determined $F = \left\langle
  |\hat{\sigma}(2\pi/L,0)|^2+|\hat{\sigma}(0,2\pi/L)|^2\right\rangle/2$
and obtained the correlation length via~\cite{ballesteros2001} via
\begin{equation}\label{eq:xi}
 \xi  \equiv  \frac{1}{2\sin(\pi/L)}\sqrt{\frac{\langle M^2\rangle}{F}-1}.
\end{equation}
To estimate the limiting value of $\xi / L$ we relied
on the quotients method~\cite{fytasRFIM,night,bal96}: The temperature where
$\xi_{2L} / \xi_{L} = 2$, \emph{i.e.}, where the curves of $\xi / L$ for the sizes
$L$ and $2L$ cross (see the inset of Fig.~\ref{fig:cor_length}) defines the finite-size pseudo-critical points. Let us denote the value of $\xi / L$ at these crossing points
as $(\xi/L)^{\ast}$. In the main panel of Fig.~\ref{fig:cor_length} we show $(\xi/L)^{\ast}$ for
all the three values of $\Delta$ and the five largest pairs of system sizes as listed in the caption of this figure. The solid lines are a joint polynomial fit, third order in $L^{-\omega}$~\cite{quotients}, where as usual $\omega = 7/4$, with a shared infinite-limit size extrapolation, leading to
\begin{equation}\label{eq:ratio_high_T}
  \left(\frac{\xi}{L}\right)_{\infty, \; \rm random} = 0.904(3).
\end{equation}
This value is beyond doubt consistent to that of the 2D Ising universality class, see Eq.~(\ref{eq:ratio-exact}) above. 

\begin{figure}
\includegraphics[width=95mm]{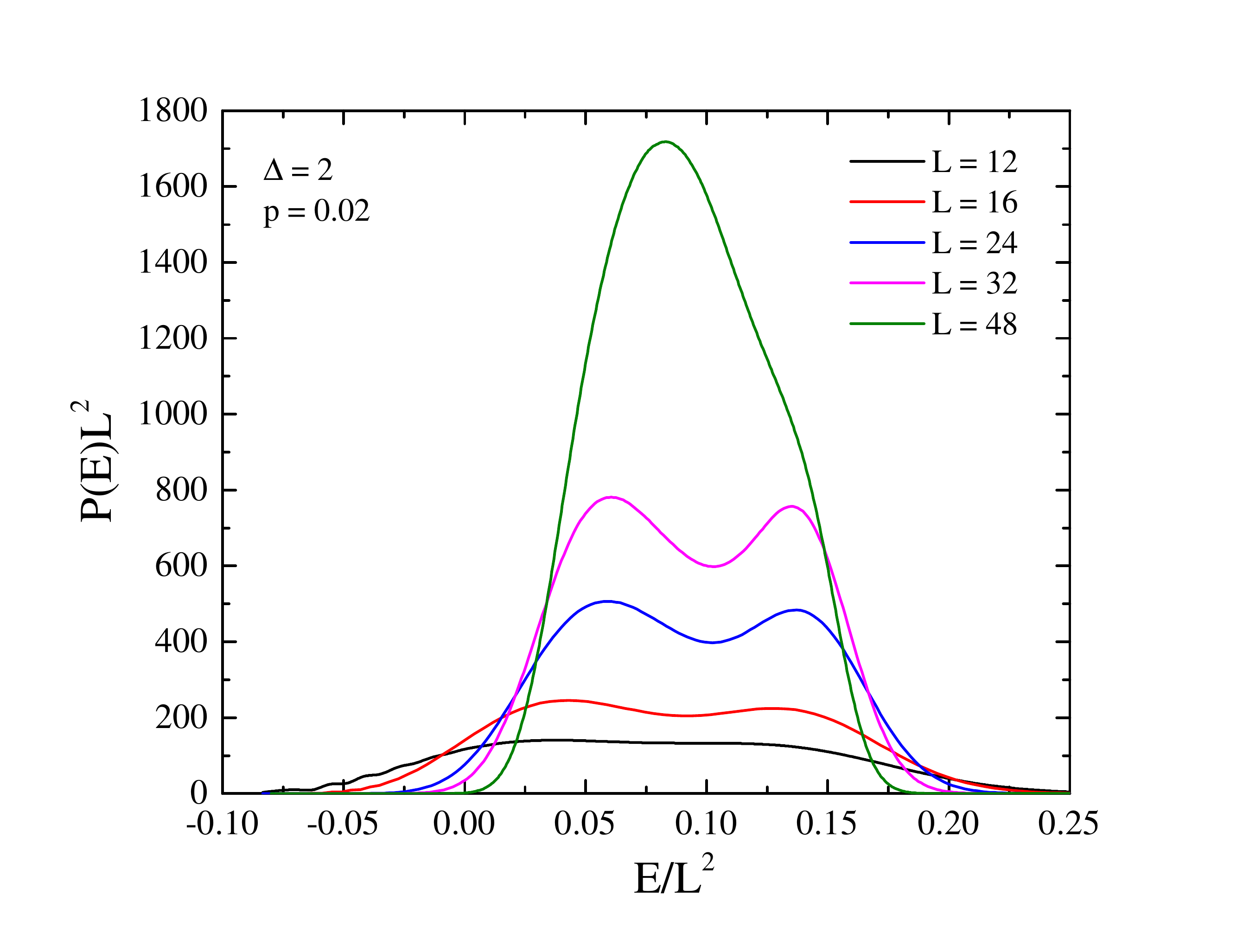}
\caption{Energy probability density functions for the full range of system sizes studied at $\Delta = 2$ and for  $p=0.02$. As in Fig.~\ref{fig:energy_pdf_L16}, the curves have been computed at the pseudocritical points of the specific heat. 
  \label{fig:energy_pdf_p002}}
\end{figure}

In the final part of our work we try to elucidate the effect of disorder on the first-order transition regime of the pure model which as discussed above can be addressed in transitions occurring in the small $p$-limit. We performed a qualitative analysis of this part of the phase diagram by fixing the crystal-field value to $\Delta = 2$ and varying the control parameter $p$, from $p=0.1$, to $p=0.06$, and finally to $p=0.02$. We simulated moderate systems with linear sizes up to $L = 48$ and averaged over $200$ samples of the randomness distribution. The results for the magnetic susceptibility are shown in Fig.~\ref{fig:chi_small_p} for the series of system sizes studied as outlined in the panel. There are three sets of curves corresponding to the three values of $p$, as indicated. A clear shift to smaller pseudocritical temperatures is observed as $p\rightarrow 0$. In fact we should note the following values for the pseudocritical temperatures of the largest system size considered ($L = 48$): $T^{\ast}_{48}(p=0.1) \approx 0.88$, $T^{\ast}_{48}(p=0.06) \approx 0.76$, and $T^{\ast}_{48}(p=0.02)\approx 0.58$. We would like to remind the reader that the tricritical point of the pure model is located at the temperature $T_{\rm } \approx 0.608$~\cite{kwak15} (see also Fig.~\ref{fig:phase_diagram}), indicating that for the case $\Delta = 2$ the value $p=0.02$ would correspond to the ex-first-order transition regime of the model. Moreover, whereas for the cases $p=0.1$ and $p=0.06$ the curves appear to be indicative of a continuous transition, for $p=0.02$ there is a pronounced increase of the peak which may be reminiscent of a first-order transition. Accordingly in the inset of Fig.~\ref{fig:chi_small_p} we depict the specific-heat data for the case of interest $p=0.02$ where similar conclusions may be drawn. 

\begin{figure}
\includegraphics[width=95mm]{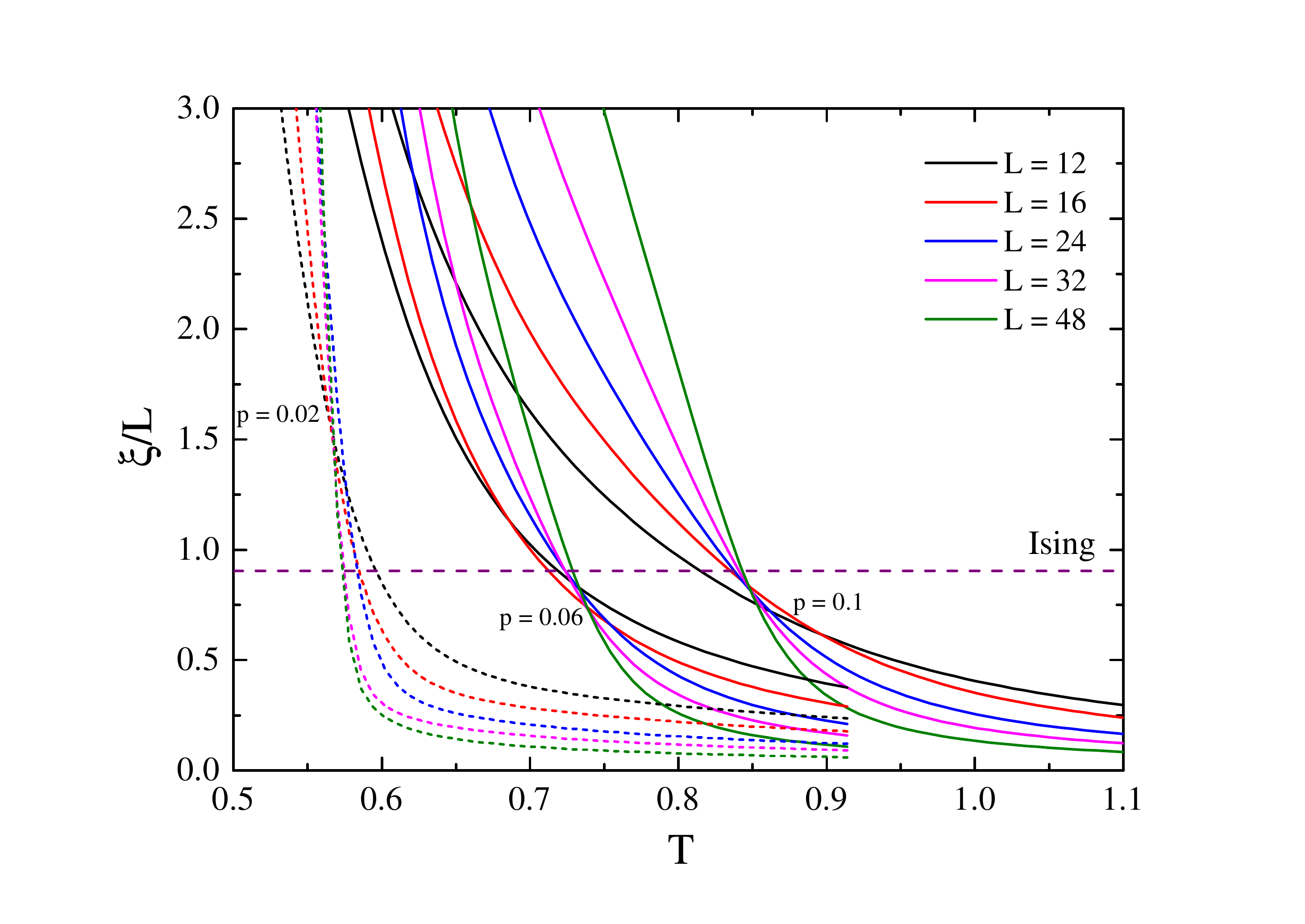}
\caption{Typical $\xi/L$ curves as a function of the temperature for various system sizes for the case $\Delta = 2$ and for $p=0.1$, $0.06$ (solid lines) and $p=0.02$ (dashed lines). The temperature area of the crossings conforms to the values identified in Fig.~\ref{fig:chi_small_p} from the maxima of the susceptibility. The The horizontal dashed line shows the asymptotic value for the square-lattice Ising model with periodic boundaries, see Eq.~(\ref{eq:ratio-exact}). 
  \label{fig:xi_small_p}}
\end{figure}

To corroborate these results in a more systematic way we present in Figs.~\ref{fig:energy_pdf_L16} and \ref{fig:energy_pdf_p002} some typical energy probability density functions at the small $p$-limit. In particular, in Fig.~\ref{fig:energy_pdf_L16} the first-order nature of the transition at $\Delta = 2$ is manifested as $p\rightarrow 0$ for a system with linear size $L=16$. Note the single-peak structure of the energy density for the cases $p=0.1$ and $p=0.06$ in comparison to the double-peak structure for the case with $p=0.02$, characteristic of a first-order transition. However, as it may be seen in Fig.~\ref{fig:energy_pdf_p002}, where the case $p=0.02$ is examined in detail, this double-peak structure is a mere finite-size effect that prevails in the regime of small to moderate system sizes. Clearly, with increasing system size the energy density exhibits only a single, symmetric peak, illustrating the second-order nature of the transition in the limit $L\rightarrow \infty$ and therefore the expected smoothening effect of disorder. This is clear evidence that the randomness distribution~(\ref{eq:bimodal}) for $p=0.02$ changes the pure first-order phase transition at $\Delta = 2$ into a disorder-induced continuous one, yet, with a crossover behavior for small system sizes. This crossover length appears to be of the order of $L^{\ast} \approx 48$ which should be taken as the minimum size in any  finite-size scaling analysis of the model at this regime of the phase diagram.

Although such an analysis goes beyond the scope of the present work, some instructive conclusions can already be drawn simply by inspecting the finite-size scaling behavior of the correlation length at the $p\leq 0.1$ regime, see Fig.~\ref{fig:xi_small_p}. For the cases with $p=0.1$ and $0.06$ the $(\xi/L)^{\ast}$ values at the crossing points approach the universal value of the Ising ferromagnet, as it should be expected, and with rather small scaling corrections. Similar results have been presented in Ref.~\cite{fytas15} for the two-dimensional random-bond 8-states Potts model, where the randomness-induced continuous transition was also shown to belong to the universality class of the pure 2D Ising ferromagnet. However, the data for the case $p=0.02$ are affected by strong scaling corrections [note that for the pair $(L,2L) = (24,48)$, $(\xi/L)^{\ast} \approx 2.5$] and this is in agreement with the existence of the crossover length discussed above. We expect that for this particular case the size evolution is non-monotonic and the true asymptotic behavior will settle in for sizes $L\gg 48$ (similar observations have been made in Ref.~\cite{fytas18} for the random-bond version of the square-lattice Blume-Capel model).

\section{Summary and Outlook}
\label{sec:conclusions}

\begin{figure}
\includegraphics[width=95mm]{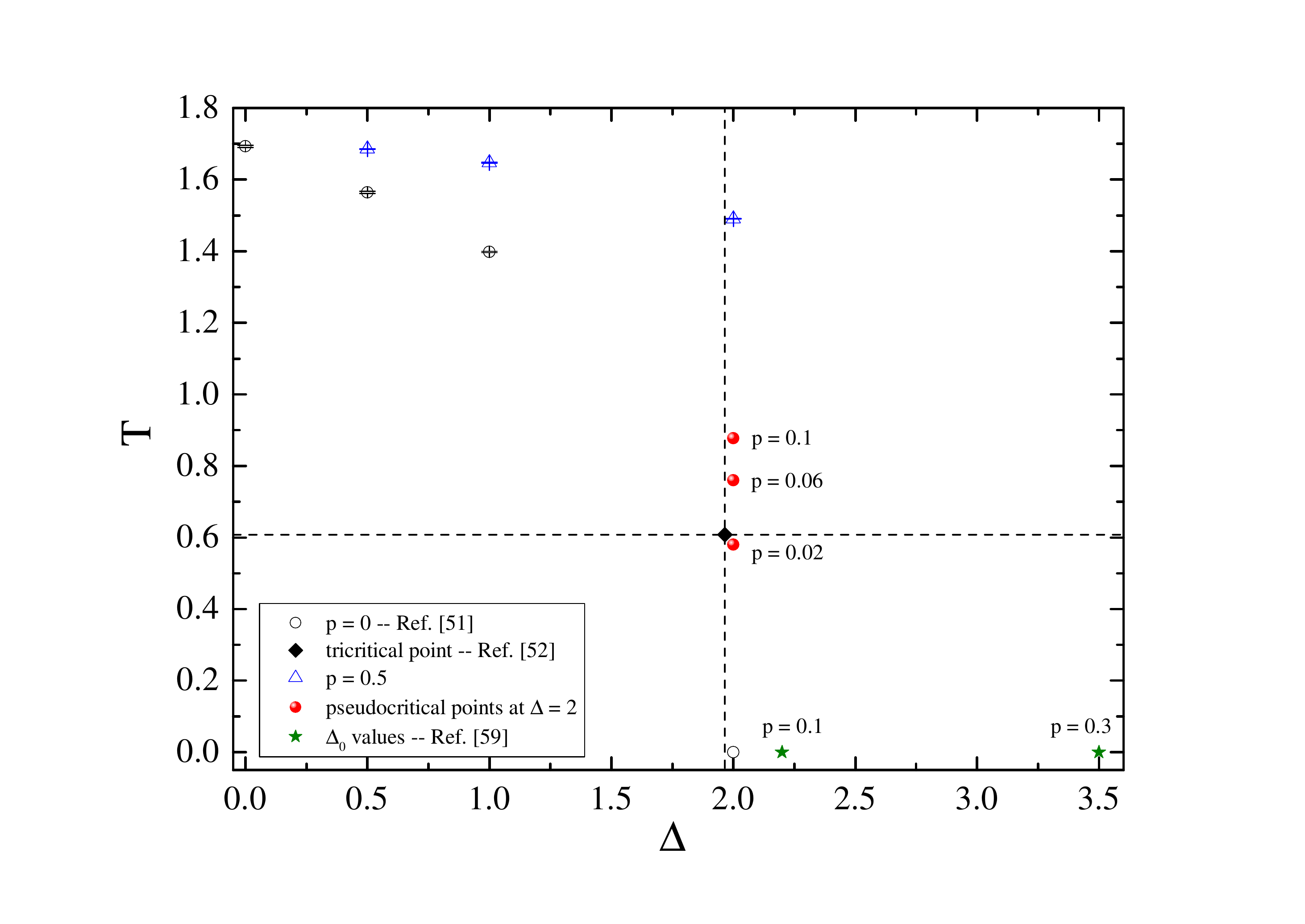}
\caption{Selected critical and pseudocritical points of the pure and disordered Blume-Capel model as defined in Eqs.~(\ref{eq:Ham_dis}) and (\ref{eq:bimodal}).
  \label{fig:phase_diagram_random}}
\end{figure}

We have studied the two-dimensional Blume-Capel model with a quenched random crystal-field coupling. On the numerical side, our work demonstrated that a parallel implementation of the Wang-Landau algorithm based on distributed memory architectures is an asset in the study of magnetic spin systems under the presence of quenched disorder. On the physical side, our results can be summarized in Fig.~\ref{fig:phase_diagram_random} upon which we shall base our concluding remarks. In this plot we have added several data points from the current work but also from previous works to allow for an overview of the current understanding of the system. The black open circles are selected critical points of the pure ($p = 0$) model taken from Ref.~\cite{malakis09} for $\Delta = 0$, $0.5$, and $\Delta = 1$ where the model falls into the universality class of the pure Ising ferromagnet. Note the value $\Delta_{0}(p=0) = 2$. The black filled rhombus (with the accompanying horizontal and vertical dashed lines) depicts the tricritical point of the pure model, taken from Ref.~\cite{kwak15}. The blue open triangles are critical points for the random $p=0.5$ model at $\Delta = 0.5$, $1$, and $2$ estimated in the current work. For these critical points Ising universality has been established in terms of critical exponents and the universal ratio $\xi/L$ with the addition of logarithmic corrections at the scaling of the specific heat. A simple argument has already been given above in Sec.~\ref{sec:results} that explains the increase of the critical temperature of the random model in comparison to that of the pure model, a behavior which is in agreement with the renormalization-group results of Ref.~\cite{branco97}. The red filled circles are pseudocritical points that correspond to the maxima of the susceptibility of the size $L= 48$ of the disordered system at $\Delta = 2$ and for $p=0.1$, $0.06$, and $p=0.02$  as obtained in the current work. These results indicate that in order to probe signatures of the first-order transition regime of the model but also to identify the smoothening effects of the disorder one should focus on the small $p$-limit and in fact the set of parameters $(\Delta = 2, p = 0.02)$ may be a promising choice. For this case, strong finite-size effects and crossover phenomena make their appearance and obscure the application of finite-size scaling. However, we do believe that the softened continuous transition will also belong to the Ising universality class, but very large system sizes will be needed in order to arrive at a safe conclusion. Finally, the green filled stars are estimates of $\Delta_{0}$ for $p=0.1$ and $p=0.3$ taken from Ref.~\cite{branco97} that agree nicely with the overall trend of the critical points for the different values of the control parameter $p$.

To conclude, although universality is a cornerstone in the theory of critical phenomena, it stands on a less solid foundation for systems subject to quenched disorder. An explicit confirmation of the behavior of disordered models in this respect is
therefore of fundamental importance for the theory as a whole (see also
Ref.~\cite{fytasRFIM}). We hope that the findings of the present work will trigger
additional studies of similar systems (i) in two dimensions, where for weak disorder
the appearance of crossover phenomena is unavoidable and their dependence on the randomness parameters remains uncharted~\cite{fytas18,tomita01}, and (ii) more importantly in three dimensions, where randomness is only relevant beyond a finite threshold~\cite{hui:89a,berker93,chatelain:01a,chatelain:05}. This still unsettled field of research alongside with a dedicated study of the effects of random-fields on the critical behavior of the three-dimensional Blume-Capel model are some of the main topics that we would like to pursue in the near future. Finally, it might be also interesting to investigate the effect of randomness exactly at the tricritical point of the two-dimensional model. Perhaps there would be just an effect on the width of the crossover region, \emph{e.g.}, changing from a double-logarithmic behavior to a simple-logarithmic one in the specific-heat scaling. In fact, a similar behavior has been observed upon introducing a very weak randomness in the Ising case~\cite{wang:90}. Finally, it is worth noting that several useful results on the implications of the scaling theory for the crossover phenomena in disordered systems close to a tricritical point can be found in Refs.~\cite{binder92,eichhorn95}.

\begin{acknowledgments}
N.G.F. is grateful to W. Selke and N.S. Branco for their communication and some very 
useful suggestions. We would also like to thank two anonymous referees
for their constructive comments. The numerical calculations reported in this paper were
performed at T\"{U}B\.{I}TAK ULAKB\.{I}M (Turkish agency), High Performance and
Grid Computing Center (TRUBA Resources). This research was supported in part by PLGrid Infrastructure.
\end{acknowledgments}

{}

\begin{thebibliography}{199}
\bibitem{quotients} Within the framework of the quotients method and for dimensionless magnitudes, say $g$, such as the Binder cumulant and the ratio $\xi/L$, a scaling ansatz of the form $g^{\ast}_{L} = g_{\infty} + \mathcal{O}(L^{-\omega})$ is expected, where $g_{\infty}$ and the corrections-to-scaling exponent
$\omega$ are universal.
\end{thebibliography}
\end{document}